\documentstyle[aps,amsfonts,bm,12pt]{revtex}

\newcommand{\bra}[1]{\mbox{$\langle{#1}|$}}
\newcommand{\ket}[1]{\mbox{$|{#1}\rangle$}}
\newcommand{\diracsp}[2]{\mbox{$\langle{#1}|{#2}\rangle$}}
\def\I{{\rm i}}
\def\d{{\rm d}}
\def\e{{\rm e}}

\begin{document}
\draft
\title{Construction of a Scalable, Uniform and Universal\\
Quantum Network and 
Its Applications
\thanks{Supported by the National Natural Science Foundation of China under 
Grant No. 69773052}}

\author{An Min WANG$^{1,2,3}$}
\address{CCAST(World Laboratory) P.O.Box 8730, Beijing 100080, People's 
Republic of China$^1$\\
and Laboratory of Quantum Communication and Quantum Computing\\
University of Science and Technology of China$^2$\\
Department of Modern Physics, University of Science and Technology of China\\
P.O. Box 4, Hefei 230027, People's Republic of China$^3$}


\maketitle

\vspace{0.1in}

\begin{abstract}
We present a possible candidate of construction of a scalable, uniform and universal quantum network, which is built from quantum gates to elements of quantum circuit, again to quantum subnetworks and finally to an entire quantum network. Our scheme can overcome some difficulties of the existing schemes and makes improvements to different extent in the scale of quantum network, ability of computation, implementation of engineering, efficiency of quantum network, universality, compatibility, design principle, programmability, fault tolerance, error control, industrialization and commercialization {\it et. al} aspects. As the applications of this construction scheme, we obtain the entire quantum networks for Shor's algorithm, Grover's algorithm and solving Schr\"odinger equation in general. This implies that the  scalable, uniform and universal quantum networks are able to generally describe the known main results and can be further applied to more interesting examples in quantum computations. 
\medskip

\end{abstract}

\section{Introduction \label{S1}}

The combination of information science and quantum mechanics has created 
a series of amazing results. One of them is just the idea of quantum 
computer \cite{Feynman,Deutsch} which can speed up a computation 
greatly and even exponentially than a classical computer, 
for example, factorization of a large number \cite{Shor} and search 
for the unstructured data \cite{Grover}. The rapid developments both 
in theory and experiment seem to indicate that the quantum computer is very promising. 

A central part of the quantum computer is a quantum network (quantum circuit) without including initial input device and final output device. The quantum network consists of wires and gates, but where the wires carry qubits, and the gates are unitary transformations. Moreover, a quantum network is an array of quantum gates, in which the quantum gates are assembled and connected according to some principles and rules. The quantum computer that can execute just one generic two-qubit gate is adequate to perform a universal quantum computation -- it can approximate any unitary transformation acting on $n$ qubits to any desired accuracy. In other words, a universal quantum network can be constructed by a set of elementary gates in principle \cite{Barenco}. 

Note that the universality in the known scheme of construction of quantum network \cite{Barenco} refers to the construction method of quantum networks, but not the computation abilities and applications of quantum networks. That is, a unitary transformation (quantum computation) always is able to be decomposed to an array of quantum gates in mathematical \cite{Reck}, and so the method to construct the quantum networks in terms of quantum gates is universal. However, such decompositions heavily depend on the forms of unitary transformations. This implies that in the known scheme, the quantum networks for different quantum computations generally have different structures of array of quantum gates, that is different structures of quantum hardwares. Thus, a given quantum network in the known scheme is only designed and made for a particular quantum computation and can not be used to the other quantum computations in general at the hardware level. Even if this given quantum network can be programmed so that it can carry out the other quantum computations, this corresponding programming language is obliged to design particularly. Consequently, a given quantum network in known scheme is generally special at software level. In the above senses, the quantum networks in the known scheme seem not to be really universal for its abilities and applications. (Note the meaning of universality). Based on the knowledge and experiences in the classical computer, a scalable and uniform structure is generally an essential precondition of universality for its abilities and applications. However, comparing with the importance of fault tolerance, error control and efficiency of quantum network, the scalable and uniform structure of quantum networks is secondary. For the purposes to improve fault tolerance, strengthen error control and increase efficiency of quantum network, we indeed can give up the requirements of the scalable and uniform structure of quantum networks. Nevertheless, the known scheme of quantum networks is not so. The reason that the known scheme of quantum networks depending on the different quantum computations has not considered the problems about the scalable and uniform structure is not to guarantee fault tolerance, error control and efficiency of quantum networks in spite of the efficiency of error correction techniques had ever been analyzed (for example \cite{Cirac,Miquel}). We think that further developing and improving known scheme or finding new schemes are very important and interesting for practical applications with the development of quantum computers. Of course, any new construction scheme of quantum networks considering the scalable, uniform structure and universality can not pay the price to depress abilities of fault tolerance, error control and efficiency of quantum networks in general. 

Although the known scheme \cite{Barenco} has become an actual standard in the design of quantum network and resulted in some beautiful and significant achievements, it seems to us that there exist some difficulties and shortcoming within it. For example, one only constructed an effective quantum subnetwork for a part of quantum Fourier transformation acting on a given state, but had not obtained its entire quantum network. Again, one did not know how to directly construct an entire quantum network to solve Schr\"odinger equation (simulating quantum dynamics) in terms of the method of discretization time. This is because the method directly by use of the total time evolution operator to construct a quantum network for solving Schr\"odinger equation leads to departing from the physical idea of discretization time. Two difficulties are original from the fact that the existing schemes can not naturally, simply and effectively construct an entire quantum network by assembling and connecting its parts -- quantum subnetworks. The deeper reasons, it seems to us, are also that the existing schemes of quantum networks have not the scalable and uniform structures.  

Therefore, in this paper, we would like to present a possible candidate of construction of a scalable, uniform and universal quantum network. The quantum networks in our scheme are built from quantum gates to elements of quantum circuit, again to quantum subnetworks and finally to the entire quantum networks. Our scheme can overcome some difficulties of the existing schemes and makes improvements to different extent in the scale of quantum network, ability of computation, implementation of engineering, efficiency of quantum network, universality, compatibility, design principle, programmability, fault tolerance, error control, industrialization and commercialization {\it et. al} aspects, which will be summarized in the last of this paper. As the applications of this construction scheme, we obtain the entire quantum networks for Shor's algorithm, Grover's algorithm and solving Schr\"odinger equation in general. This implies that the scalable, uniform and universal quantum networks are able to generally describe the known main results and can be further applied to more interesting examples in quantum computations. 

Our construction scheme is not to put the known quantum gate-assembly schemes into a nicety and concretization because our ideas have been beyond the known schemes. Not only the construction of a salable, uniform and universal quantum network, but also {\it how to effectively construct} and {\it how to implement a quantum network in experiments} are very important. In this paper, we will consider how to obtain a construction with high efficiency in principle by using of our method. However, we do not intend to directly touch at the great difficulties on the experimental implement. 

Our main propose in this paper is to obtain a standardized and uniform way to construct an entire quantum network and universally describe a general quantum computation. It must be emphasized that we do not try to find a physical system to implement a concrete quantum network here, but we only study how to construct a scalable, uniform and universal quantum network from a theoretical view.  

It is ture that the existing schemes of quantum networks have achieved great success,  but our believe is ``let's try every thing" just as the fact that one are trying to the different implements of quantum gates (or quantum computational elements) in experiments with individual advantages and features. Maybe one of them is better within a period, this does not implies that the others have not their vitality. More possible candidates of construction of quantum networks will be able to provide us more methods and choices to build practical quantum computers in near future. 

The arrangement of the paper is as follows. In Sec. \ref{S2}, we account for our more motivations to propose the new possible candidate of construction of a scalable, uniform and universal quantum network and define some basic terms frequently used in this paper. In Sec. \ref{S3}, two main difficulties in the known quantum gate-assembly schemes are pointed out in order to give the concrete reasons why we present our scheme. In Sec. \ref{S4}, Our construction scheme for a scalable, uniform and universal quantum network is presented from a theoretical view. The main features of our scheme are given out. In Sec. \ref{S5}, we give out the entire quantum network for Shor's and Grover's algorithms. In Sec. \ref{S6}, we obtain an entire quantum network for generally solving Schr\"odinger equation. In the last, we summary that our scheme makes improvements over the existing schemes to different extent in the scale of quantum network, ability of computation, implementation of engineering, efficiency of quantum network, universality, compatibility, design principle, programmability, fault tolerance, error control, industrialization and commercialization {\it et. al} aspects we concern with. 

\section{More motivations and basic concepts \label{S2}}

It must be emphasized that in the known quantum network models, one has assumed that qubits can be initialized in a particular standard state and measured in a particular standard basis, the final output of a quantum computation is obtained by measuring the qubits, and the quantum network always corresponds to a unitary transformation (or a sequence of unitary transformations to affect simultaneously). 
It seems to us, the third assume, that a quantum network always corresponding a unitary transformation, can restrict the scale of a quantum computation. Just as is well known, in a classical algorithm, in order to increase programming and computing efficiency, one often needs circulation and conditional choose {\em et. al}, that is that the intermediate results have to be used. We think that the case is so in a quantum algorithm. Thus, to read "intermediate quantum states" has to introduce quantum measurements unless the intermediate quantum states can be distributed to known address and this address can be correctly arrived at, or storage the intermediate quantum states in auxiliary memory. A quantum measurement must breaks the unitary property, and distributing a unknown quantum state or use much auxiliary memory must pay the price to decrease efficiency and then is not economy. This is the first reason. Our second reason is that the decoherence time of a quantum system may be so short that a quantum computing task can not be finished during it. One needs to obtain the intermediate results before the decoherence of this system and then carries out further quantum computation. Quantum measurement will be introduced for this purpose in general. This means that the unitary property can not be kept in the whole process. The last, a quantum computation can be speeded up in terms of quantum entanglement, but the product of quantum entanglement of the whole system is very difficult at present. However, by use of quantum entanglement of subsystems that is respectively easier to product, one can obtain the needed intermediate results and then based them one implements further quantum computation. Again, quantum measurement is needed and the unitary property is gone. From these three reasons, we prefer to think that in our scheme, a quantum computing task consists of, in general, some computing steps and or computing parts as well as quantum measurements, every step and part not including quantum measurements corresponds to a unitary transformation. In fact, the existing quantum algorithm are in general so. 

In the senses stated above, the known quantum network model is only suitable to construct a respectively ``smaller scale" quantum network or a subnetwork of a quantum computing task. However, for many practical quantum computing tasks to scale quantum networks or to assemble quantum subnetworks is necessary. This problem can be extended to how to construct an entire quantum network in terms of quantum gates or some elementary components of quantum circuit so that this entire quantum network can carry out a general quantum computing task. It is clear that the known scheme has not given how to do and the new schemes are needed. 

Actually, it is often difficult to directly decompose a quantum computation with somewhat large scale to the elementary gates \cite{Reck}. This is because the mathematical symmetries and physical features of a total quantum computation often are very complicated, even can not be directly found out as well as utilized. Thus the known scheme to construct quantum networks is too complicated to easily do, even is not feasible in practical. However, it must be respectively easier and often can be done to finds the partial mathematical symmetries and physical features of such some parts of quantum systems, for example, quantum Fourier transformation. Sometimes,  decomposing directly a quantum computation to the elementary gates may be conflict with the well-known physical idea. For instance, solving Schr\"odinger equation in general can be based on a discretization of the time evolution in a series of short enough interval \cite{Wiesner,Zalka}. However, constructing a quantum network directly for a total transformation will back to a finite time and lose the advantage of discretization time. This can not do it in physics. Therefore, to construct the subnetworks of the parts and steps of quantum computation and then to connect and assemble them together to form an entire quantum network is just a requirement from practical applications. 

At present, although there are several excellent quantum algorithms, 
but the knowledge about constructions of their entire quantum networks 
are still not completed. Moreover, the quantum networks in the known scheme have not simple, scalable and uniform structures. These shortcoming will increase the difficulty of design quantum networks, even depress the efficiency of quantum networks, and finally block the development of quantum computer. 

Obviously, any possible candidates of construction of quantum networks had better to be compatible with the existing schemes so that the results and achievements from the existing schemes can be inherited. This problem has been considered in our construction scheme. 

Just based on the motivations stated above, we present our construction scheme of a scalable, uniform and universal quantum network. Our start point is that a quantum computation usually consists of a series of quantum computing steps or a set of quantum computing parts unless it is very simple and particular. It is very natural to think that an entire (or total) quantum network is made of some quantum subnetworks in which each of them corresponds to a quantum computing step including quantum measurement or a quantum computing part. 

By the word quantum subnetwork, we mean a quantum network corresponding 
to a quantum computing step or a quantum computing part in a quantum 
computation. Suppose a quantum computation $U$ is expressed by a product of a series of quantum computing steps, that is $U=U_1U_2\cdots U_n$, or a summation 
of a set of quantum computing parts, that is $U=U_1+U_2+\cdots+U_n$. Thus, 
a quantum network corresponding to $U_i (i=1,2,\cdots,n)$ is a quantum 
subnetwork with respect to the entire (or total) quantum network for $U$. A quantum measurement also can be thought of a quantum computing step (but non-unitary). 
Moreover, a quantum subnetwork can have its hiberarchy more than one level. 
The number of its hiberarchy depends on the requirement how to decompose 
an entire quantum network so that the constructions for the lowest 
hierarchy of the quantum subnetworks are effective and facile enough 
as possible. A quantum sub-subnetwork is always called with respect 
to its upper hierarchy quantum subnetwork. 

In order to connect and assemble quantum subnetworks easily together or build an entire quantum network in terms of its quantum subnetworks, we introduce an auxiliary qubit (or a quantum system) and successfully find a realized method. Then, we obtain   that a quantum subnetwork can be constructed by directly connecting a series of the 
elements of quantum circuit together. The elements of 
quantum circuit have two kinds of basic elements respectively called as ``Rotator" and ``Transitor", and two kinds of the auxiliary elements respectively called as ``Jointer" and ``Connector" , which will be introduced in Sec. \ref{S4}.
A element of quantum circuit, such as the rotator and transistor, is clearly defined by the matrix elements in a transformation matrix corresponding to a quantum computing step or a quantum computing part. On the other hand, an element of 
quantum circuit also can be  made of the elementary quantum gates. 
Consequently, the procedure in our construction scheme is fully 
determined. That is, we present a possible candidate of construction of a scalable, uniform and universal quantum network, which is built from quantum gates to elements of quantum circuit, again to quantum subnetworks and finally to an entire quantum network.

Recently, there is a series of new results in experiments for implementing 
quantum computation. One of the reasons is that every possible 
thing is being tried. For example, Ahn {\it et al.} did not use $n$ qubits as a recourse, but rather a single atom prepared in a superposition of $n$ Rydberg state \cite{Knight}. In contrast, without starting with quantum gates, but introducing elements of quantum circuit as the essential components of quantum (sub-)network is also a possible way to construct quantum network. Although our elements of 
quantum circuit can be constructed by quantum gates, 
perhaps, their experimental implements do not need to be given through combining the implements for quantum gates, but can directly obtained by suitable quantum systems. If this guess comes true, then the elements of quantum circuit will be more fundamental in a quantum network. 
  
\section{Difficulties in known quantum gate-assembly schemes\label{S3}}

The first example is quantum Fourier transformation. It plays an important 
role in quantum computation including factorization, search algorithms and 
quantum simulating. Its matrix $F$ reads
\begin{equation}
F=\frac{1}{\sqrt{N}}\sum_{m,n=0}^{N-1}\e^{2\pi\I mn/N}\ket{x_m}\bra{p_n},
\quad F^{-1}=\frac{1}{\sqrt{N}}\sum_{m,n=0}^{N-1}\e^{-2\pi\I mn/N}
\ket{p_m}\bra{x_n}.
\end{equation}
Quantum Fourier transformation can be rewritten \cite{Ekert} as
\begin{equation}
F=\sum_{n=0}^{2^k-1}\prod^k_{j=1}\frac{1}{\sqrt{2}}(\ket{0}+\e^{2^j\I\pi 
n/(2^k-1)}\ket{1})\bra{p_n}=\sum_{n=0}^{2^k-1}\prod_{\otimes,j=1}^k B_j
(2^j\pi  n/(2^k-1)H_j\ket{0}\bra{p_n}.
\end{equation}  
where $B_j$ is a rotator gate and $H_j$ is a Hadamard gate acting on the 
$j-$qubit. Every term in the above summation can be regarded as the 
construction for a quantum (sub-)network with high efficiency in the 
known schemes. However, The total transformation is not an ideal form of 
a quantum network since it is not a product of its quantum 
subnetworks. In other words, one only constructed the effective quantum 
subnetwork for a part of quantum Fourier transformation acting on a given 
state $\ket{p_n}$, but has not obtained its entire quantum network. 
Obviously, 
this is not enough, for example, Shor's algorithm needs an entire quantum 
network for quantum Fourier transformation. Therefore, it is necessary to 
find such a method so that an entire quantum network can easily be built by 
its quantum subnetworks. If this is done, then we not only can utilize the 
known some effective constructions, but also we can have a powerful mean to 
seek an effective construction, that is, to decompose a quantum network 
into such a set of the quantum subnetworks that they can be effectively 
constructed as possible.   
   
The second example is about solving Schr\"odinger 
equation in general or simulating quantum dynamics. In quantum mechanics, the evolution of Schr\"odinger wavefunction 
with time can be written as:
\begin{equation}
\psi(x,t)=T\exp\left\{-\I \int_{0}^t \d\tau H\right\}\psi(x,0),
\end{equation}
where $T$ is a time-order operator and the natural unit system is taken. 
When $H$ does not obviously depend on time, it is simplified as
\begin{equation}
\psi(x,t)=\e^{-\I Ht}\psi(x,0).
\end{equation}
Choosing a short time step $\triangle t$, the time evolution operator 
$\e^{-\I Ht}$ reads
\begin{equation}
\Omega(\triangle t)=(1-iH\triangle t).
\end{equation}
If we wish to advance the system by time $T$, we can repeat the whole 
process $T/\triangle t$ times, that is
\begin{equation}
\Omega(T)=(1-iH\triangle t)^{T/\triangle t}.
\end{equation}
Now we would like to construct a quantum network for the time evolution 
operator according to the known quantum gate-assembly schemes. But, there 
exists an obvious difficulty. If we start directly from the total time 
evolution operator, we will lose the advantage of the discretization time to solve Schr\"odinger equation. Actually it seems not to be feasible since departing from the known-well physical idea. A very natural idea is to construct a quantum subnetwork 
for $(1-iH\triangle t)$ and then connect all such quantum 
subnetworks together. However, the known quantum gate-assembly scheme 
has not provided how to do it. This, again, call us to find a method which 
can easily build an entire quantum network in terms of its quantum subnetworks.

\section{Construction in theory \label{S4}}

\subsection{Universal quantum networks with uniform structure}

A quantum computation, in mathematical, can be described by a (unitary) transformation matrix $U$ acting on an input state of a quantum system, 
that is
\begin{equation}
\ket{\Psi}^\prime=U\ket{\Psi}=\sum_{m,n} U_{mn}\ket{m}\bra{n}\sum_{s}
\psi_s\ket{s}=
\sum_{m,n}U_{mn}\psi_n\ket{m},
\end{equation}
and each component of this quantum state is transformed as
\begin{equation}
\psi_m^\prime=\sum_{n=0}^{2^k-1}U_{mn}\psi_n. \label{QCT}
\end{equation}
The output state $\ket{\Psi}^\prime$ contains the needed information. 
Based on our ideas stated above, we first introduce an auxiliary qubit $A$ 
prepared in $\ket{0}_A$ and define two kinds of basic elements of quantum 
circuit $R_m$, $T_{mn}$ 
\begin{eqnarray}
R_m(U_{mm})&=&\exp\{(U_{mm}\ket{m}\bra{m}\otimes I_A)\cdot C^\dagger\}=
I_R\otimes I_A+(U_{mm}\ket{m}\bra{m}\otimes I_A)\cdot C^\dagger,\\
T_{mn}
(U_{mn})(m\neq n)
&=&\exp\{(U_{mn}\ket{m}\bra{n}\otimes I_A)\cdot C^\dagger\}=I_R\otimes I_A+(U_{mn}\ket{m}\bra{n}\otimes I_A)\cdot C^\dagger,
\end{eqnarray}
where $I_R$ and $I_A$ are respectively unit matrices in the register space and the 
auxiliary qubit space, $N_A$ is a NOT gate acting on the 
auxiliary qubit. While $C^\dagger$ is an auxiliary element of quantum 
circuit
\begin{equation}
C^\dagger=I_R\otimes\ket{1}{}_A{}_A\bra{0}=I_R\otimes c_A^\dagger,
\end{equation}
which satisfies that $C^{\dagger\;2}=0$. This leads to an exponential 
form to reduce to only two terms. It is clear that the action of 
$C^\dagger$ is able to joint the transformations for various basic 
vectors, that is the different elements of quantum circuit together. 
For example, $R_m(\alpha)R_n(\beta)=I_R\otimes I_A+[(\alpha\ket{m}\bra{m}+\beta\ket{n}\bra{n})\otimes I_A]\cdot C^\dagger=\exp\{(\alpha\ket{m}\bra{m}+\beta\ket{n}\bra{n})\otimes I_A]\cdot C^\dagger\}$.  So, $C^\dagger$ can be called as a``Jointer". Furthermore, 
since $c_A=\ket{0}{}_A{}_A\bra{1}$ and $c_A^\dagger=\ket{1}{}_A{}_A\bra{0}$, 
it is easy to verify that $c_A^2=c_A^{\dagger 2}=0; c_A c_A^\dagger+
c_A^\dagger c_A=I_A$.  
Thus $c_A$ and $c_A^\dagger$ can be thought of the fermionic annihilate 
and create operator respectively in the auxiliary qubit. From this, the 
auxiliary system might be able to be extended to a larger system beyond a 
qubit if one can implement easily a nilpotent transiting operator. 

$R_m(U_{mm})$ can be called the ``{\it Rotator}" for its action makes  $\ket{m}\otimes\ket{0}_A$ to rotate to $\ket{m}\otimes\ket{0}_A+U_{mm}
\ket{m}\otimes \ket{1}_A)$. 
$T_{mn}$ can be called the ``{\it Transitor}" for its action makes $\ket{n}\otimes\ket{0}_A$ to map to $\ket{n}\otimes\ket{0}_A+U_{mn}\ket{m}\otimes\ket{1}_A$. 

It is worth emphasizing that there is an essential difference between 
a classical gate and a quantum gate. It is just that a classical gate 
is always to carry out a determined operation, but a quantum gate can 
carry out a kind of operations. For example, a quantum rotation gate 
can rotate the state to any angle and then it needs a parameter $\phi$ 
or $\e^{\I\phi}$ to determine its operations. So do the rotator and 
transistor. Each rotator $R_{n}$ or transitor $T_{mn}$ depends on one 
parameter $U_{mm}$ or $U_{mn}$.  

Note that the elements of quantum circuit are not obviously unitary, one would like to know how to implement them in experiments. Actually, the unitary property of the elements of quantum circuit is broken by quantum measurement. We will see that they can be written as the products of a series of unitary transformations and quantum measurement in the following part of this section. So they can be implemented in experiments in principle. In quantum mechanics, the elements of quantum circuit correspond to the reversible transitions between two states and thus they are able be obtained in experiments.
    
Since the action of the jointers, the rotator and transitor can be 
connected together and form a $(k+1)-$qubit quantum network, 
\begin{eqnarray}
Q(U)
&=&\prod_{m,n=0}^{k-1}\exp\{(U_{mn}\ket{m}\bra{n} \otimes I_A) 
C^\dagger\}\\
&= &\prod_{m,n=0}^{k-1}\exp\{[(U_{mn}\ket{m}\bra{n} +U_{nm}
\ket{n}\bra{m})\otimes I_A/2]\cdot C^\dagger\}. \label{UQN}
\end{eqnarray}
where $k$ qubits form the register space and a qubit belongs to the 
auxiliary space.  It is easy to obtain its reversible operation
\begin{eqnarray}
Q^{-1}(U)
&=&\prod_{m,n=0}^{k-1}\exp\{-(U_{mn}\ket{m}\bra{n} \otimes I_A) 
C^\dagger\}\\
&= &\prod_{m,n=0}^{k-1}\exp\{-[(U_{mn}\ket{m}\bra{n} +U_{nm}
\ket{n}\bra{m})\otimes I_A/2]\cdot C^\dagger\}. \label{IUQN}
\end{eqnarray}

Therefore, $Q(U)$ acting on $\ket{\Psi}\otimes\ket{0}_A$ just carries out a 
general quantum computation which is reversible:
\begin{equation}
Q(U)\ket{\Psi}\otimes \ket{0}_A=\ket{\Psi}\otimes \ket{0}_A+
\ket{\Psi}^\prime\otimes \ket{1}_A.
\end{equation}
Moreover, two project measurements 
\begin{equation}
D=C^\dagger C=I_R\otimes\ket{1}{}_A{}_A\bra{1},\quad P=C C^\dagger=I_R\otimes\ket{0}{}_A{}_A\bra{0}
\end{equation}
result in an output state $\ket{\Psi}^\prime\otimes\ket{1}_A$ and  
an input state $\ket{\Psi}\otimes \ket{0}_A$ respectively in a quantum 
computation (\ref{QCT}). 

Obviously, our construction scheme of quantum network is uniform. In fact,  every element of quantum circuit is like a ``building block". They are simply putting 
together even without the order limitation, and all of them form a quantum 
(sub-)network. 

\subsection{Scalable quantum networks}

It must be emphasized that the action of the jointer plays an important role 
for construction of a scalable, uniform and universal quantum network. Because of the jointers, we can obtain a quantum network for a summation 
of a set of quantum computing parts
\begin{equation}
Q(U)=Q(U_1+U_2+\cdots+U_r)=Q(U_1)Q(U_2)\cdots Q(U_r)\label{QNforS},
\end{equation}
where $Q(U_i)$ is a quantum subnetwork for a quantum computing part $U_i$. 
This is one of the most important and new features of our construction. 

Introducing a so-called ``Connector"  defined by  
\begin{equation}
C=I_R\otimes \ket{0}{}_A{}_A\bra{1}=I_R\otimes c_A\label{connector},
\end{equation}
which is used to prepare a intermediate state so that this prepared state can 
be used in a series of successive transformations, we can obtain a quantum 
network for a product of a series of quantum computing steps 
\begin{eqnarray}
Q(U)&=&Q(U_1U_2\cdots U_r)= I_R\otimes I_A+ C^\dagger\left(\prod_{j=1}^{r} CQ(U_j)\right) CC^\dagger \\
&=&I_R\otimes I_A+ \tilde{Q}(U)=\exp\{\tilde{Q}(U)\}\label{QNforP}, 
\end{eqnarray} 
where $Q(U_i)$ is a quantum subnetwork for a quantum computing step $U_i$. 
This is also one of the most important and new feature of our construction.
 
Note that in Eq.(\ref{QNforP}) $I_R\otimes I_A$ is added so that the 
transformation $U_1U_2\cdots U_r$ is reversible. The another way is to 
use two registers respectively to input state and out state and the latter 
includes one auxiliary qubit. Thus, the quantum network for product of a 
series of quantum computing steps becomes a form of full multiplication:
\begin{equation}
\bar{Q}(U)=\bar{Q}(U_1U_2\cdots U_r)=(I_R)_{\rm in}\otimes \left[C^\dagger\left(\prod_{j=1}^{r} CQ(U_j)\right) CC^\dagger\right]_{\rm out}=(I_R)_{\rm in}\otimes \tilde{Q}(U)\label{QNforPP},
\end{equation}
while the initial state is now prepared as $(\ket{\Psi(t)})_{\rm in}
\otimes (\ket{\Psi(t)}\otimes \ket{0}_A)_{\rm out}$. In special, if 
$U=U_1\otimes U_2\otimes\cdots\otimes U_r$, we have
\begin{eqnarray}
\bar{Q}(U)&=&\bar{Q}(U_1\otimes U_2\otimes\cdots\otimes U_r) \nonumber\\
&=&(I_R)_{\rm in}\otimes \left[C^\dagger\left(\prod_{j=1}^{r} CQ(I_1\otimes\cdots\otimes I_{j-1}\otimes U_j\otimes I_{j+1}\otimes\cdots
\otimes I_r)\right) CC^\dagger\right]_{\rm out} \nonumber\\
\label{QNforDPP}
&=&(I_R)_{\rm in}\otimes \tilde{Q}(U)\label{QNforPPP},
\end{eqnarray}
Furthermore, we can obtain
\begin{equation}
\bar{Q}((U_1\otimes U_2\otimes\cdots\otimes U_r)V)=(I_R)_{\rm in}\otimes 
C^\dagger C\tilde{Q}(U_1\otimes U_2\otimes\cdots\otimes U_r)CQ(V)CC^\dagger,
\end{equation}
This means that the role of $\tilde{Q}$ is the same as one of $Q$ when we 
connect them by the connectors. But, the role of $\tilde{Q}$ is different 
from one of $Q$ when they are the quantum networks for a summation of 
various quantum computing parts. Actually, this is a reason why we take a 
quantum subnetwork for a set of quantum computing parts to be, in general, 
a lower hierarchy than a quantum subnetworks for a series of quantum 
computing steps. 

Therefore, our scheme of quantum network is scalable in terms of the above two important features. An entire quantum network is then constructed {\it via.} a so-called ``motherboard" with many ``slots". A slot is a interspace between a pair of connectors and then a motherboard is made of some connectors. The quantum subnetworks for a series of quantum computing steps, just like some ``plug-in boards", are inserted into these slots, and the quantum subnetworks for a set of quantum computing parts, even like some ``building blocks", are put together directly. All of them are so easily assembled and connected as an entire quantum network which can carry out a 
general quantum computation. 

\subsection{Elements of quantum circuit and quantum gate}

Now, let's build the relations between the elements of quantum circuit and 
the quantum gates. In mathematics, the elements of quantum 
circuit are the natural basis of the operator in Hilbert space and each 
of them directly relates with a matrix element of the transformation 
matrix for a quantum computation. Of course, they can be constructed by the elementary quantum gates \cite{Shor2}. It is easy to see
\begin{eqnarray}
\ket{m}\bra{n}
&=&\frac{1}{2^k}\prod_{\otimes, i=0}^{k-1}[\delta_{\alpha_i 0} \delta_{\beta_i0}(I+Z)+\delta_{\alpha_i 1}\delta_{\beta_i1}(I-Z)\\ \nonumber
& &+\delta_{\alpha_i 0}\delta_{\beta_i1}(X+Y)+\delta_{\alpha_i 1}
\delta_{\beta_i0}(X-Y)];\\ 
\ket{m}&=&\prod_{\otimes, i=0}^{k-1}\ket{\alpha_i},\qquad \bra{n}=
\prod_{\otimes, i=0}^{k-1}\bra{\beta_i},
\end{eqnarray} 
where $X=\sigma_x; \I Y=\sigma_y; Z=\sigma_z$ and $\sigma_{x,y,z}$ are 
usual Pauli spin matrix. In fact, $\ket{m}\bra{n}$ can be written as a 
product of a general exchange transformation between two adjacent states 
and a measurement $\ket{n}\bra{n}$ from left or a measurement 
$\ket{m}\bra{m}$ from right. To do this, we introduce the generalized 
exchange gate $E(m,m+1)$ for two adjacent basic states $\ket{m}$ and 
$\ket{m+1}$ defined by
\begin{equation}
E(m,m+1)=E(m+1,m)=\sum_{j=0;j\neq m,m+1}^{2^k-1}\ket{j}\bra{j}+\ket{m}\bra{m+1}+\ket{m+1}\bra{m}.
\end{equation}
Obviously, it is Hermian and unitary. It is a two state gate but not a 
qubit gate in general. It is easy to see that it acts $\ket{m}$ or 
$\ket{m+1}$ leads to their exchange and keeps the other basic states 
invariant. Note that for two qubits, $E(2,3)$ is a CNOT gate and $E(1,2)$ 
is a swapping gate \cite{Loss}. The exchange transformation of arbitrary 
two states $\ket{m}$ and $\ket{n}$ can be constructed by 
\begin{equation}
E(m,n)=\left\{
\begin{array}{ll}
\displaystyle\prod_{j=n}^{m-1}E(j+1,j) &\quad (n<m),\\
\displaystyle\prod_{j=0}^{n-m-1}E(n-j-1,n-j) &\quad (n>m),\\
I_R &\quad (n=m).
\end{array}\right.
\end{equation}
The order of product is arranged from the left side with index $j$ 
increasing. Therefore, $E(m,n)(m\neq n)$ can be expressed by a product of $|m-n|$ successive the generalized exchange gates in which each of 
the generalized exchange gate is unitary and only involves two adjacent 
states. In special, $E(m,m)$ is just an identity gate. Obviously, $E(m,n)\ket{n}=\ket{m}$ and $\bra{m}E(m,n)=\bra{n}$. This implies that 
$E(m,n)$ is a transiting unit from $\ket{n}$ to $\ket{m}$. So an 
arbitrary transformation $U$ for $k$ qubits can be written as 
\begin{equation}
U=\sum_{m,n=0}^{2^k-1}U_{mn}E(m,n)\ket{n}\bra{n}=\sum_{m,n=0}^{2^k-1}U_{mn}\ket{m}\bra{m}E(m,n).
\end{equation}
Thus the scalable, uniform and universal quantum network for $U$ is just defined by
\begin{equation}
Q(U)=\prod_{m,n=0}^{2^k-1}\exp\{(U_{mn}E(m,n)\ket{n}\bra{n}\otimes I_A)
\cdot C^\dagger\}.\label{UQNE}
\end{equation}
If the graphic rules for the elements of quantum circuit with form $\exp\{(U_{mn}E(m,n)\ket{n}\bra{n}\otimes I_A)\cdot C^\dagger\}$ are given 
out, the picture of the quantum network $Q(U)$ can be drawn easily, because 
the construction of a quantum network is constructed by directly connecting 
the elements of quantum circuit together. In fact, the picture of the 
elements of quantum circuit such as rotators and transitors also can be 
drawn if we introduce the graphic rules for the general exchange gate for 
two adjacent states and the measurement gate. 

It is worth noting that an element of quantum circuit is constructed by 
a transiting unit $E$ and a measurement unit and can be written as an exponential 
form. Perhaps, it is helpful to implement an element of quantum circuit 
in experiment and understand why the elements of quantum circuit may be 
fundamental in our construction scheme. Obviously, $Q(U)$ is universal 
and the Eq.(\ref{UQNE}) is an alternative of Reck {\it et.al}'s formula 
\cite{Reck} since it keeps the advantages such as the product form, only 
involving two states (not qubit) and the closed relation with the 
elementary gates {\it et.al}. Even our construction can be applied to an 
irreversible and/or non-unitary transformation. This means we can have 
more ways to seek an effective construction and even new quantum 
algorithms.

\subsection{Efficiency of quantum network and compatibility with the known scheme}

The key point is the efficiency of quantum network. It seems 
that in our scheme a quantum network consisting of $2^k\times 2^k$ basic 
elements is the same as a classical computer by use of computing resources. 
This is a price to reach at universality. Because the basic elements 
of quantum circuit only act on a branch (path) of the quantum data flow from 
its definition, $Q(U)$ needs $2^k\times 2^k$ basic elements of quantum circuit in general. However, in practice, our construction is obtained in terms of the symmetries and physical features in a quantum computation. This leads that many elements of quantum circuit do not really appear. For example, the simplest is a 
quantum network for a transformation of a single quantum state: $\ket{n}\longrightarrow \e^{\I\alpha_n}\ket{n}, \ket{m}(m\neq n)\longrightarrow\ket{m}$
\begin{equation}
Q(S(\e^{\I\alpha_n}))=\exp\{C^\dagger\}\exp\{[(\e^{\I\alpha_n}-1)\ket{n}\bra{n}\otimes I_A]\cdot C^\dagger\}
\end{equation}
where $\exp\{C^\dagger\}=Q(I_R)$ 
is a quantum subnetwork for an identity gate and we construct 
$Q(S(\e^{\I\alpha_n}))$ by decomposing this transformation into 
$I+(\e^{\I\alpha_n}-1)\ket{n}\bra{n}$. In special, when $\alpha_n=\pi$, 
this transformation is a reflection of $\ket{n}$, which can be used in 
Grover's algorithm. Another example is a quantum subnetwork 
\begin{equation}
Q(U(i))=\exp\{[I_1\otimes I_2\otimes\cdots\otimes U(i)\otimes\cdots\otimes I_k)\otimes I_A]\cdot C^\dagger\}
\end{equation}
for a transformation only acting on the $i-$th qubit which leads $\ket{\alpha_1}\otimes\ket{\alpha_2}\otimes\cdots\otimes\ket{\alpha_i}
\otimes\cdots\otimes\ket{\alpha_k}$ to $\ket{\alpha_1}\otimes\ket{\alpha_2}\otimes\cdots\otimes(U(i)\ket{\alpha_i})
\otimes\cdots\otimes\ket{\alpha_k}$. In special, for a controlled gate 
$U_{\rm control}=\ket{0}\bra{0}\otimes I+\ket{1}\bra{1}\otimes U$, we have
\begin{equation}
Q(U_{\rm control})=\exp\{(\ket{0}\bra{0}\otimes I_2\otimes I_A)C^\dagger\}\exp\{(\ket{1}\bra{1}\otimes U\otimes I_A) C^\dagger\}
\end{equation}
When $U$ is a NOT gate, it is a quantum network for a controlled NOT. 
Likewise, we can write down a quantum network for Toffoli gate (controlled-controlled-NOT), in which the third qubit experiences NOT if and only if 
the others are in the state $\ket{1}$: 
\begin{equation}
Q({\rm Toffoli})=\exp\{C^\dagger\}\exp\{-(\ket{1}\bra{1}\otimes
\ket{1}\bra{1}\otimes I\otimes I_A)C^\dagger\}\exp\{(\ket{1}\bra{1}\otimes\ket{1}\bra{1}\otimes N\otimes I_A)C^\dagger\}
\end{equation}
since we have the decomposition $U_{\rm Toffoli}=I-\ket{1}\bra{1}\otimes
\ket{1}\bra{1}\otimes I+\ket{1}\bra{1}\otimes\ket{1}\bra{1}\otimes N$. 
Therefore, all the elementary quantum gates can be written down by our scheme of 
quantum networks. This implies that our construction scheme is compatible 
with the known quantum gate-assembly schemes. As soon as we have 
obtained an expression of a quantum computation in terms of a set of the 
elementary quantum gates, we can first construct the quantum subnetworks 
for these elementary quantum gates and then assemble easily them into a 
entire quantum network by virtue of the properties of our quantum network. 
Therfore, the efficiency of quantum network in our scheme is not lower than the one in the known scheme.

It must be emphasized that our construction has further simplification 
and more ways to build a universal quantum network. One of the simplest 
examples is a quantum network for a transformation of a single state 
stated above. Actually, for a diagonal transformation, its quantum network 
can be defined by $Q(U_d)=\prod_{m=0}^{k-1}\exp\{(U_{m}\ket{m}\bra{m}\otimes I_A)\cdot C^\dagger\}$. Moreover, Rotator, Transitor, even their combinations with many branches can be introduced. For example, the conditional rotation gate $R_2$ only acting on the second qubit in 3-qubit register is made of two Rotators 
with $2^3/2$ branches. \cite{Ekert} Another typical case is the computing step $V$ can be written as the direct product $V(1)\otimes V(2)$. Suppose $V(1)$ is $2^{k_1}\times 2^{k_1}$ and $V(2)$ is $2^{k_2}\times 2^{k_2}$ $(k_1+k_2=k)$. Then, 
if in $V(1)$ we can decrease a parameter or an elements, the result 
leads that $2^{k_2}\times 2^{k_2}$ parameters or the number of the 
corresponding elements are decreased. The use of computing resource is 
then at high efficiency. So, to simplify $U$ into the direct product of 
subspaces as possible is a better method to increase the efficiency of the 
use of computing resources. Quantum Fourier transformation only acting on a given state is just an example. Its quantum network can be effectively constructed since the above reason \cite{Ekert}. From the definition of our scalable, uniform and universal quantum network 
Eq.(\ref{UQN}) and its property Eq.(\ref{QNforS}), it is easy to get an 
entire quantum network for a quantum Fourier transformation in terms of 
its quantum subnetworks \cite{Ekert} 
\begin{equation}
Q(F)=\prod_{n=0}^{2^k-1}Q[B(n)HM_{0n}]= \prod_{m=0}^{2^k-1}\prod_{n=0}^{2^k-1}\exp\{[(B(n)H)_{m0}\ket{m}\bra{n}\otimes I_A] \cdot C^\dagger\},
\end{equation}
where $B(n)H=\prod_{\otimes,j=0}^{2^k-1}B_j[2^j\pi n/(2^k-1)]H_j$ and $M_{0n}=\ket{0}\bra{x_n}$. Because, each quantum subnetwork for a part of 
Fourier transformation acting on a given state is an effective construction, 
the entire quantum network $Q(F)$ is also so. The difficulty in the known 
construction scheme of quantum network is then overcome. 

Our scheme provides not only a procedure to simplify the construction of an entire  
quantum network, but also a way to obtain an effective construction. 
That is, we can decompose a total quantum computation into a summation of 
a set of quantum computing parts and/or a product of a series of quantum 
computing steps so that quantum subnetworks for them can be constructed 
effectively as possible. Note that a quantum subnetwork for a set of 
quantum computing parts is usually a lower hierarchy than a quantum 
subnetworks for a series of quantum computing steps. As soon as we obtain 
the effective constructions for the quantum subnetworks, an effective 
construction for an entire quantum network is just direct and facile. 

Generally speaking, the number of the elements of quantum circuit in a quantum network for the transformation $U$ is at least equal to the number of the different matrix elements in $U$ expect for zero. In general, the principle to decompose a quantum computing task into a summation of a set of quantum computing parts and/or a product of a series of quantum computing steps is that the symmetries and physical features in these quantum computing parts or steps can be found easily as possible. The symmetries and physical features in a quantum computing part, step or a whole computation will largely decreased the number of the element of quantum circuit. Thus, in principle, we can finally obtain an effective construction for a quantum network. 

\subsection{The other features}

It is very important and interesting how to program a quantum network 
\cite{Nielsen}. In our construction, the transformation parameters in a 
quantum computation are directly and explicitly related with the elements 
of quantum circuit. Resetting them is just program a quantum network. 
Therefore, our construction for a universal quantum network is potentially programmable.

It necessary to point out that there are three important features of the 
connector introduced by our construction. First, a connector is a standard 
interface unit between quantum subnetworks. In a classical algorithm, one 
usually does not need to consider how to connect two computing steps because 
the classical data flow is generally single branch (unless in parallel), 
but, in a quantum algorithm, one has to think over this problem because 
the quantum data flow is generally many branches, which can be arranged in different forms. If with respect to two quantum computing steps or parts, that is two unitary transformations, one 
designs their quantum networks independently, then the arrangement ways of 
quantum data of input and output for two quantum subnetworks are different 
in general. This means that an interface unit is needed. How to design this 
interface unit just becomes a problem. Here, in our construction scheme for a quantum network, connector as a standard interface unit is 
introduced and one does not worry about this problem again. Second, the 
interspace between two connectors looks like a ``slot" for a quantum 
subnetwork corresponding to a quantum computing step and the quantum subnetwork appears as a ``pinboard". For example, 
the quantum network for quantum Fourier transformation can be reused in a 
entire quantum network for Shor's algorithm or Grover's algorithm by 
plugging in the corresponding slot. At present, one still does not know 
how to program a quantum network. This means that a quantum network is 
only able to carry out a given quantum computation. Therefore, the quantum subnetworks is made as plug-in and reusable one is useful and economy, because many resources can be significantly saved. 
Finally, the connector provides a tool to assemble and scale some quantum 
(sub-)networks in order to carry out a larger quantum computation.  

From this scalable, uniform and universal quantum network proposed by this paper, it follows a general design principle for a quantum algorithm and quantum simulating. 
That is, we ought to find such a suitable and optimized decomposition 
that the quantum subnetworks for its components can be effectively and 
easily constructed as possible. The process of decomposition can continue 
until our aim is arrived at. In principle, this is possible if a quantum 
computation can be effectively carried out. In order to do this, we need 
to use the fundamental laws of physics, specially the principles and 
features of quantum mechanics, for example, coordinate system choice, 
representation transformation, picture scheme and quantum measurement 
theory, if we have thought that a quantum computing task is a physical 
process. Moreover, we have to use the symmetry properties of every 
computing parts and/or step $U^i$ as possible, such as the direct product decomposition, the transposing invariance $U_{nm}^i=U_{mn}^i$ or the row 
equality $U_{mn}^i=U_{m0}^i$ for all $n$ or the line equality $U_{mn}^i=
U_{0n}^i$ for all $m$ as well as make $U_i$ with zero elements and equal 
elements as many as possible. In general, we always can find such some 
symmetries through decomposing $U$ into a summation $U_1+U_2+\cdots+U_n$ 
and/or a product $U_1U_2\cdots U_n$. 

In the above senses, we present the construction scheme of a scalable, uniform and universal quantum network which can carry out a general quantum computing. Our construction scheme can keep the advantages in the known schemes, can overcome some difficulties of the known schemes, are compatible with the known schemes, and has new features easy to scale and with uniform structure. Moreover, each step in constructing procedure of our quantum network is clear, determined and facile. Because there are many new features, our scheme makes improvements to different extent in the scale of quantum network, ability of computation, implementation of engineering, efficiency of quantum network, universality, compatibility, design principle, programmability, fault tolerance, error control, industrialization and commercialization {\it et. al} aspects, which are summarized in the end of this paper.

\section{Applications to quantum algorithm \label{S5}}

In this section, we would like to show that the entire quantum networks of 
the known main quantum algorithms can be described by our scalable, uniform and universal quantum network. 

The most famous example is Shor's algorithm which is used to the factorization of a 
large number $N$. It can exponentially speed up the computation in a quantum 
computer than in a classical computer. Shor's algorithm can be 
described by the following five steps. First, 
start with two $k-$qubit registers in $\ket{0}_1\ket{0}_2$, then prepare the 
first register into a superposition state with the equal weight in terms of 
Fourier transformation or $k-$qubit Hadamard gate denoted by $H$:
\begin{equation}
H\ket{0}_1\ket{0}_2=\sum_{n=0}^{2^k-1}\ket{n}_1\ket{0}_2.
\end{equation}
Second, select randomly a factor $a$ and make a mapping
\begin{equation}
\sum_{n=0}^{2^k-1}\ket{n}_1\ket{0}_2\stackrel{G}{\longrightarrow} \sum_{n=0}^{2^k-1}\ket{n}_1\ket{a^n {\rm mod} N}_2.
\end{equation}
Obviously
\begin{equation}
G=\sum_{n=0}^{2^k-1}\ket{n}_1\ket{a^n {\rm mod}N}_2\bra{n}_1\bra{0}_2.
\end{equation}
Third, measure the second register by $I_1\otimes \ket{a^m {\rm mod}N}
\bra{a^m {\rm mod}N}$ and obtain the result: $\sum_{j=0}^{[2^k/r]-1} |jr 
+ l\rangle |u\rangle$. 
Fourthly, do Fourier transformation $F$ to the first register so that
\begin{equation}
U_{\rm DFT}\ket{jr+l}_1= \frac{1}{\sqrt{2^{k}}}\sum_{y=0}^{2^k-1} \exp 
\{2\pi {\rm i}(jr+l)y/2^k\} \ket{y}_1.
\end{equation}
and obtain the final state $\displaystyle \frac{1}{\sqrt{r}}\sum_{m=0}^{r-1} 
\exp (2\pi {\rm i}lm/r) |m2^k/r\rangle $. The last, measure the first 
register in the basis $y=m2^k/r$. If one obtains one values $y$, then 
solve equation $y/2^k=m/r$ to find the period. Once $r$ is known the 
factors of $N$ are obtained by calculating the greatest common divisor 
of $N$ and $a^{r/2}\pm 1$. 

Thus, the main steps in Shor's algorithm can be represented by one total transformation matrix:
\begin{equation}
U({\rm Shor})= (F\otimes I_2)M(a^m {\rm mod}N)GH;\quad M(n)=I_1\otimes\ket{n}_2\bra{n}_2.
\end{equation}

The product of the serval matrices is an easy problem. After we know the 
form of $U({\rm Shor})$, we are able to obtain all of the parameters, 
that is the elements of $U$, to determine the construction of a quantum 
network for Shor's algorithm in our method. Since the third step is a quantum measurement, it breaks the unitary property of $U$ as a total quantum computation task. In order to construct the entire quantum network in terms of Barenco {\it et.al}'s method in principle, usually one respectively constructs the quantum network for modular exponentiation and Hadamard transformation \cite{Vedral1,Miquel}, and quantum network for quantum Fourier transformation \cite{Ekert}. However, how to connect them and how to scale the quantum Fourier transformation for a given state to the a superposition state have not been considered. Therefore, the existing quantum network for Shor's algorithm is still not completed. On other hand, one proves the quantum measurement can commute with quantum Fourier transformation and then the quantum network for Shor's algorithm also can be constructed in principle by Barenco {\it et. al}'s method. However, the methods of two kinds of constructions, the known scheme and our scheme, are directly done can not guarantee their construction is effective in general. Actually, the most important significance of Shor's algorithm is based on the fact that he found an effective decomposition of quantum algorithm of factorization to five quantum computing steps. Obviously, starting from this decomposition one can more easier construct an effective quantum network. One of our scheme's aims is just so. In other words, the key skills in our scheme are to find an suitable and optimal decomposition of a quantum computing task, then design the effective constructions of quantum subnetworks for every components in this decomposition, and finally assemble and connect these quantum subnetworks to build an entire quantum network. 

According the principle stated above, we use the decomposition given by Shor and design the effective quantum subnetworks for every step. Since we have obtained an effective construction of the quantum network for quantum Fourier transformation in Sec. \ref{S5}, we only need to find the quantum subnetworks for $H$ and $G$. Obviously, starting from the initial state $\ket{0}_1\otimes\ket{0}_2\otimes\ket{0}_A$, it follows that $Q(H)$ is
\begin{equation}       
\bar{Q}(H)=I_{\rm in}\otimes C^\dagger\left(\prod_{j=1}^k (CQ(H_j\otimes I_2))\right) C C^\dagger=I_{\rm in}\otimes \tilde{Q}(H).
\end{equation}
Note that $H_j$ is a Hadamard transformation only acting on the $j$-qubit 
in the first register and the second register has kept the original state. 
While the realization of our quantum network for mapping $G$ can read
\begin{equation}
Q(G)=\prod_{n=0}^{2^k-1}\exp\{(\ket{n}\ket{a^n {\rm mod}N}\bra{n}\bra{0})
\otimes I_A)\cdot C^\dagger\}.
\end{equation}
Furthermore, in terms of the connectors, the entire quantum network for 
Shor's factorization can be obtained as the following:
\begin{equation}
\bar{Q}({\rm Shor})=(I_R)_{\rm in}\otimes\left(C^\dagger CQ(F\otimes I_2)
M(a^m{\rm mod}N)\otimes I_AC Q(G)C \tilde{Q}(H) CC^\dagger\right)_{\rm out}.
\end{equation} 
Obviously, this entire quantum network for Shor's algorithm is an effective 
construction because that its main quantum subnetworks for Fourier 
transformation, modular exponentiation and Hadamard transformation are effective. 

In order to consider the losses and decoherence in a factoring computer, we can use  Miquel {\it et.al}'s result \cite{Miquel}, which the efficiency of some simple error correction techniques were analyzed. That is, before the third step, we can use Vedral \cite{Vedral1}, Miquel {\it et.al} designed quantum network $U_{\rm VM}$ for modular exponentiation and Hadamard transformation as a subnetwork. Thus, the entire quantum network for Shor's factorization also can be obtained as the following:
\begin{equation}
{Q}_{h}({\rm Shor})=(I_R)_{\rm in}\otimes\left(C^\dagger CQ(F\otimes I_2)
M(a^m{\rm mod}N)\otimes I_A CC^\dagger\right)_{\rm out} (I_R)_{\rm in}\otimes (U_{\rm VM}\otimes I_A)_{\rm out}.
\end{equation} 
The initial state is prepared as $(\ket{0}_1\otimes\ket{0}_2)_{\rm in}\otimes (\ket{0}_1\otimes\ket{0}_2\otimes\ket{0}_A)_{\rm out}$. Note that in general the heterotic connection of quantum networks with different structures needs an interface unit. However, in our scheme we have set the input state as a standard and simple form and so we can extend $U_{\rm VM}$ to $(I_R)_{\rm in}\otimes (U_{\rm VM}\otimes I_A)_{\rm out}$ to realize the connection without an additional interface unit.  

The heterotic connection between our quantum subnetworks and the known quantum subnetworks not only displays the compatibility between our scheme and the existing schemes, but also implies that we have more means, including to use the known results, to implement error control.  
 
Another famous quantum algorithm is Grover's algorithm which is used to 
search the expected term in an unstructured data. It can be described by 
the following four steps. First, start with a $k-$qubit registers in 
$\ket{0}$, then prepare it into a superposition with the equal weight 
in terms of Fourier transformation or $k-$Hadamard gate, that is $H\ket{0}=\sum_{n=0}^{2^k-1}\ket{n}$. 
Second, do a reflection:
\begin{equation}
R_2=I-2\ket{j}\bra{j}=\sum_{n=0}^{2^k-1}(-1)^{\delta_{jm}}\ket{x_m}\bra{x_m},
\end{equation}
where $j$ corresponds to the expected data. Third, 
make the following operation:
\begin{equation}
R_1=F^{-1}R_0F=F^{-1}[2\ket{0}\bra{0}-I]F=-F^{-1}\sum_{m=0}^{2^k-1}
(-)^{\delta_{0 m}}\ket{x_m}\bra{x_m}F,
\end{equation} 
where $I$ is an identity matrix, $F$ is a quantum Fourier transformation, 
$F^{-1}$ is its inverse and $R_0=2\ket{0}\bra{0}-I$. The last, repeat 
$R_1R_2$ $\sqrt{N}\pi/4$ times and then do all measurements. 

Since the quantum network for quantum Fourier transformation has been 
obtained and its inverse has the similar realization but its parameters 
with a negative sign. While $R_0$ and $R_2$ is diagonal, it is very easy 
to get from our construction for a quantum network
\begin{eqnarray}
Q(R_0)&=&\exp\{(2\ket{0}\bra{0}\otimes I_A)\cdot C^\dagger\}
\exp\{-C^\dagger\}\\ 
Q(R_2)&=&\exp\{C^\dagger\}\exp\{(-2\ket{x_j}\bra{x_j}\otimes I_A)
\cdot C^\dagger\}.
\end{eqnarray}
Thus the quantum network for Grover's algorithm is just obtained
\begin{equation}
Q({\rm Grover})=(I_R)_{\rm in}\otimes\left(C^\dagger CQ(F^{-1})CQ(R_0)CQ(F)CQ(R_2)C\tilde{Q}(H)CC^\dagger\right)_{\rm out}.
\end{equation}
Again, because we have used an effective construction of quantum 
subnetwork for quantum Fourier transformation, this entire quantum 
network for Grover's algorithm is an effective construction. 

In fact, the known main quantum algorithms have realized such a 
suitable decomposition so that their every computing steps and/or 
parts have some symmetries. Thus, we first can find the effective constructions 
of the quantum subnetworks for the quantum computing steps and/or parts,  
then assemble and connect these quantum subnetworks together in terms our method,  and finally obtain an entire quantum network with effective construction.  
 
\section{Applications to quantum simulating \label{S6}}

Simulating quantum systems has such a meaning that using a specially 
designed quantum system, for example quantum computer, which is called 
a simulated system, to simulate another so-called physical quantum system. 
In 1982, Richard Feynman first proposed that a quantum system would be 
more efficiently simulated by a computer based on the principles of 
quantum mechanics rather than by one based on the principles of 
classical mechanics \cite{Feynman}. This is because that the size of 
the Hilbert space grows exponentially with increase of the number of 
particles. A full quantum simulating demands the exponential resources 
on a classical computer so that it is in general intractable. Since the 
discovery by Shor of a quantum algorithm for factoring in polynomial time \cite{Shor}, there has been tremendous activity in the field of quantum 
computations and quantum simulating. For example, Lloyd has shown 
how a quantum computer is in fact an efficient quantum simulator 
\cite{Lloyd}. In addition, some the general ideas and schemes of quantum 
simulating to several special quantum systems were proposed and discussed \cite{Wiesner,Zalka,Boghosian,Abrams,Lidar,Somaroo,Milburn,Sorensen,Sornborger}. 

At present, quantum simulating mainly performs a simulation 
of the dynamics. In this section, we simulate quantum dynamics by solving Schr\"odinger equation in general, but not by finding a specially 
designed quantum system. 

If one would like to simulate a quantum system by a quantum computer, 
the first task is how to ``write" quantum state into the quantum computer, 
in other words, how to store the information of physical system -- quantum 
state in the quantum computer. 
Although quantum computer has an ability to store information increasing exponentially than a classical computer, it deals with the information as 
digital one just like a classical computer. Consequently, a basic skill is 
to discretize the wavefunction which describes the quantum state in a 
finite space.
\begin{equation}
\psi(x=a,t)\longrightarrow \psi(x_m,t)\; \left(m=\left[\frac{a}{L/N}
\right]\right).
\end{equation}
where $L/N$ is the length of $|x_m-x_{m-1}|$. 
In addition, when the quantum system is limited within a box, it ought to 
impose the periodic boundary conditions, that is 
\begin{equation}
\psi(x_{m+N},t)=\psi(x_m,t).\label{PBC}
\end{equation} 
Therefore, the wavefunction in time $t$ can be written as a vector
\begin{equation}
\ket{\Psi(t)}=\sum_{m=0}^{N-1} \psi(x_m,t)\ket{x_m},
\end{equation}
where $\ket{x_m}$ form a set of basis with the properties of orthogonality 
and completeness ($\diracsp{x_m}{x_n}=\delta_{mn}$, $\sum_{m=0}^{N-1}\ket{x_m}\bra{x_m}=1$) in $N$-dimensional Hilbert space. When this is 
extended to two particles, it follows that
\begin{equation}
\ket{\Psi(t)}=\sum_{m_1=0}^{N_1-1}\sum_{m_2=0}^{N_2-1} \psi(x_{m_1},x_{m_2},t)\ket{x_{m_1}x_{m_2}}.
\end{equation}
Of course, the extension to high dimensional is similar. In fact, the 
above equation also can describe two dimensional case. (Usually taking 
$N_1=N_2$, this means that the particle moves in a square box).  
A quantum register with $k$ qubits can express a state in Hilbert space 
at most with $N=2^k$ dimensional. For two particles in one dimensional, 
its Hilbert space should be $N_1\times N_2=2^{k_1+k_2}$. Thus, in a 
classical computer, we need exponentially increasing bits to deal with 
the quantum state. But in a quantum computer, we see that $k_1+k_2$ qubits, 
or two quantum registers with $k_1$ and $k_2$ qubits respectively, are 
suitable to this task. In fact, this is just one of reasons why simulating 
a quantum system can be more rapidly done by use of quantum computer than 
by use of classical computer. Generally speaking, for 3-dimensional and $n$ particles, we can use $3n$ quantum registers to store the discretization the wavefunction. For the system of identical particles the initial state of the 
quantum computer has to be chosen symmetrically or anti-symmetrically. For 
quantum field theory, its discretization method can be similar to one in 
lattice gauge theory. 

It is worth emphasing how to realize the fundamental operators such as 
coordinates and momentum is a key point. In coordinate representation, 
coordinates are directly written as a diagonal matrix whose diagonal 
elements are $x_m$. But the momentum has a little complication because 
it is a derivative action as the following
\begin{equation}
\hat{p}\psi(x,t)=-\I\frac{\partial \psi(x,t)}{\partial x}=-\I
\frac{\psi(x+\triangle x,t)-\psi(x)}{\triangle x}.
\end{equation}
So the momentum operator can be defined by:
\begin{equation}
\hat{p}=-\I\;\frac{1}{2}\left(\frac{N}{L}\right)\sum_{m=0}^{N-1}(\ket{x_m}\bra{x_{m+1}}-\ket{x_m}\bra{x_{m-1}}),
\end{equation}
where, in order to make the momentum operator is Hermian,  an average 
of the left and right derivative has been taken and the periodic 
boundary conditions (\ref{PBC}) has been used. It is easy to verify
\begin{equation}
\hat{p}\ket{\Psi(t)}=\sum_{m=0}^{N-1}\left[-\I\;\frac{1}{2}
\frac{\psi(x_{m+1},t)-\psi(x_{m-1},t)}{(L/N)}\right]\ket{x_m}.
\end{equation}

Further, the kinetic energy operator (natural unit system $\hbar=c=1$) 
can be obtained:
\begin{equation}
\hat{T}=-\frac{1}{8\mu}\left(\frac{N}{L}\right)^2\left[\sum_{m=0}^{N-1}(\ket{x_m}\bra{x_{m+2}}+\ket{x_m}\bra{x_{m-2}})-2I_N\right].
\label{Tenergy}
\end{equation}

It is clear that the potential operator is a diagonal transformation 
if one only considers the local potential $U(x)$ or the external field 
$V_e(x)$, that is
\begin{eqnarray}
\hat{U}&=&\sum_{m=0}^{N-1}U(x_m)\ket{x_m}\bra{x_m},\\
\hat{V}_e&=&\sum_{m=0}^{N-1}V_e(x_m)\ket{x_m}\bra{x_m},
\end{eqnarray}        
while two-body local interaction $U(x_1,x_2)$ can be written as
\begin{equation}
\hat{U}=\sum_{m_1}^{N_1-1}\sum_{m_2}^{N_2-1}U(x_{m_1},x_{m_2})\ket{x_{m_1},x_{m_2}}\bra{x_{m_1},x_{m_2}}.
\end{equation}

It is easy to extend to high dimensional and many particles formally. 
For example, in two particles case, the fundamental momentum operators are $\hat{p}_1=\hat{p}^{(1)}\otimes I_2, \hat{p}_2=I_1\otimes \hat{p}^{(2)}$. 
Then, the kinetic energy and the potential energy operators can be 
constructed in a similar way.

Hamiltonian has been here. Now it appears, in principle, that one can use 
the known quantum gate-assembly schemes to design a quantum network for 
the time evolution operator. However, obviously if one start directly 
from the total time evolution operator, he/her will lose the advantage 
of discretization time. Actually it seems not to be feasible 
since departing from the physical idea. Even in mathematics, the quantum 
network for a time evolution operator $\Omega(t)$ in a finite time is 
very difficult to construct since $\Omega(t)$ with infinite terms. A very 
natural idea is to construct a quantum subnetwork for $(1-iH\triangle t)$ 
and then assemble all $T/\triangle t$ quantum subnetworks together. 
However, the known quantum gate-assembly scheme has not provided how to do 
it. Moreover, if one tries to write a small enough evolution operator 
within a given precision in a small time interval, then maybe one does not 
know clearly how to treat with an approximately unitary transformation in 
terms of the known schemes. In fact, the known scheme was proved only for 
an unitary transformation. In order to overcome the above difficulties, we 
use our construction scheme for a scalable, uniform and universal quantum network. First, we define the momentum quantum subnetwork as: 
\begin{eqnarray}
Q(\hat p)&=&I_R\otimes I_A+\hat p_R \cdot C^\dagger=\exp\{\hat p_R\cdot 
C^\dagger\}\\
&=&\prod_{m=1}^{2^k-1}\exp\left\{-\frac{\I}{2}\frac{L}{N}(\ket{x_m}\bra{x_m}[E(m,m+1)-E(m,m-1)]\otimes I_A)
\cdot C^\dagger\right\},
\end{eqnarray}
where 
\begin{equation}
\hat{p}_R=-\frac{\I}{2}\frac{L}{N}\sum_{m=0}^{2^k-1}\ket{x_m}\bra{x_m}
(E(m,m+1)-E(m,m-1))\otimes I_A
\end{equation}
In terms of the above method the quantum subnetworks for the kinetic 
energy and the potential energy with the factor $-\I\triangle t$ can be 
constructed as:
\begin{eqnarray}
Q(-\I\triangle tT)&=&\exp\left\{-{\I}\frac{\triangle t}{4\mu}
\frac{L^2}{N^2} C^\dagger\right\}\prod_{m=1}^{2^k-1}\exp\left\{{\I}
\frac{\triangle t}{8\mu}\left(\frac{L}{N}\right)^2
(\ket{x_m}\bra{x_m}E(m,m+2)\otimes I_A)\cdot C^\dagger\right\}\\ \nonumber 
& &\exp\left\{{\I}\frac{\triangle t}{8\mu}\left(\frac{L}{N}\right)^2(\ket{x_m}\bra{x_m}E(m,m-2)
\otimes I_A)\cdot C^\dagger\right\},
\end{eqnarray}
\begin{equation}
Q(-\I\triangle tV)=\prod_{m=0}^{2^k-1}\exp\{(-{\I}\triangle tV(x_m)\ket{x_m}\bra{x_m}\otimes I_A)\cdot C^\dagger\}.
\end{equation} 

According to the method stated above, the quantum subnetwork for the 
time evolution operator in a short time interval reads 
\begin{equation}
Q(\Omega(\triangle t))=Q(I_R)Q(-\I\triangle t H)=\exp\{C^\dagger\}
Q(-\I\triangle t T)Q(-\I\triangle t V).
\end{equation}
For a finite time, take the product of all the time evolution operators 
at small enough time steps and then obtain finally the entire quantum 
network for the time evolution operator: 
\begin{equation}
\bar{Q}(\e^{-\I Ht})= \left(I_R\right)_{\rm in}\otimes\left[C^\dagger\left(
\prod_{i=1}^{[T/\triangle t]} C Q(\Omega(\triangle t))\right)
CC^\dagger\right]_{\rm out}.
\end{equation}
By using of it, Schr\"odinger equation can be simulated in general. 
In above procedure, the entire quantum network for the time evolution 
operator is divided into $[T/\triangle t]$ quantum subnetworks for the 
time evolution operator in a short time interval, and each such quantum 
subnetwork is decomposed to two quantum sub-subnetworks $Q(T)$ and $Q(U)$ 
for the kinetic energy and potential energy respectively. Then, $Q(T)$ and 
$Q(U)$ are effectively constructed by using of the elements of quantum 
circuit. In our scheme, these quantum subnetworks are easily assembled and connected 
as an entire quantum network for solving Schr\"odinger equation 
in general.     

For two particles, the extension of this method is direct, but it is not 
efficient enough in the use of computing resources if one does it directly. 
Although the quantum network for the kinetic energy operator is obtained 
by the same method, but the quantum network for two body potential needs 
all $2^k\times 2^k$ basic elements. Thus, a better method is first to 
reduce Schr\"odinger equation to the mass center and the relative 
coordinate systems. In the mass center system, we need to simulate a 
free practice, and in the relative coordinate system, we need to simulate Schr\"odinger equation in a single-body potential action. For a free 
particle, its Hamiltonian is diagonal in the momentum representation. 
Thus, we can simulate it by a quantum subnetwork for Fourier transformation 
and a quantum subnetwork for a diagonal Hamiltonian. For the relative motion 
with a single-body potential, its quantum network is able to be obtained 
in our method stated above. This is just a example to build an effective 
construction by virtue of the physical principles. Of course, the 
efficiency problem here is also said with respect to the comparison among 
the different quantum algorithms. Because, the quantum network is built 
in quantum parallelism, that is, it acts on all states at the same time. 
Therefore, with respect to classical computing, it must be efficient. In 
this sense, the above method to solve Schr\"odinger equation can be 
extended to higher dimensional and more particles.    

\section{Improvements over the existing schemes \label{S7}}

In conclusion, our scheme makes improvements over the existing schemes to different extent in the following ten aspects that we concern with.

\begin{enumerate}
\item {\em Scale of quantum network}. Our scheme is used to construct an entire quantum network in the cases that one only knows its partial symmetry and the effective construction for its quantum subnetworks, but does not known a globe symmetry and the effective construction of the total quantum network. Quantum Fourier transformation is just an example. Our scheme also can be used to construct an entire quantum network corresponding a quantum computing task whose quantum subnetworks can   corresponds a non-unitary transformation(for quantum measurement). Our construction is very easy to assemble and scale up an entire quantum network in terms of its quantum subnetworks for the summation and product of simultaneous and successive transformations. However, this is difficult in the known quantum gate-assembly schemes. In this aspect, our construction scheme is significantly different from 
the known ones and then advances the art of construction for an entire 
quantum network. This is an important improvement. 
\item {\em Ability of computation}. Our scheme can be use to solve Schr\"odinger equation (simulating quantum dynamics) in terms of a method of discretization time, but the existing scheme seems to be unable. Our scheme also can be used to the cases when the decoherence time of a quantum system may be so short that a quantum computing task can not be finished within it. In addition, our scheme can be extended to construct quantum subnetworks for non-unitary transformations (only refer to the quantum computing steps or parts) and then can deals with more quantum computing tasks. This is an important improvement. 
\item {\em Implement of engineering}. This profits from the simple, standardized and uniform structure of quantum network in our scheme. An entire quantum network in our construction scheme is built in a ``motherboard" with many ``slots". The behavior of 
a quantum subnetwork corresponding to a quantum computing step appears a 
plug-in board in a ``slot" which is an interspace between two connectors. 
Building an entire quantum network for a product of a series of quantum 
computing steps is simply to insert such quantum subnetworks in their 
slots. The quantum (sub-)network for a summation of a set of quantum 
computing parts is even simply to put the quantum (sub-)subnetworks 
together by virtue of the so-called jointers, in which a quantum 
(sub-)subnetwork corresponding to a quantum computing part. This leads 
that a quantum network is updatable. Moreover, a quantum subnetwork and 
its any hierarchy in one quantum network can be reused in another quantum 
network. All we do is to draw it out from the slot or move it out from a 
quantum network and insert it into a given slot or put it together with 
the other subnetworks in the other quantum network. Easy to assemble, update and reuse are very important advantages in our construction scheme. However, the known quantum gate-assembly schemes are not so. This is an important improvement.     
\item {\em Universality}. At present, the existing scheme can be used to the decomposition to the array of elementary gates from a unitary transformation, but the decomposition depends on the form of the unitary transformation. For different quantum computing tasks, the construction of quantum networks in the known scheme is not really universal. However, our scheme has the universality feature with scalable and uniform structures. As the applications of our construction scheme, we have obtained the entire quantum networks for Shor's algorithm, Grover's algorithm and solving Schr\"odinger equation in general. This implies that our scalable, uniform and universal quantum network can generally describe the known main results in quantum computation. This is an important improvement.
\item{\em Compatibility} It is interesting and important whether our scheme can be used to the known decomposition from a quantum computation to a set of the 
universal quantum gates. The answer is positive, because we have proved 
that all the elementary quantum gates can be written as our quantum 
subnetworks. As soon as we know how to combine the elementary quantum 
gates to form a quantum computation, the construction of the quantum 
network in our scheme is just very easy. All that we need to do is to 
assemble the quantum subnetworks for these elementary quantum gates into 
an entire quantum network like inserting the ``plugging board" and playing 
the ``building blocks". Even we can use the known quantum network as a quantum subnetwork through an extension. For example, we can obtain a heterotic connection our quantum subnetwork for quantum Fourier transformation and Vedral and Miquel {\it et.al}'s quantum subnetwork for modular exponentiation and Hadamard transformation \cite{Vedral1,Miquel} to construct an entire quantum network for Shor's algorithm (see in Sec. \ref{S5}). Therefore, we can say that our construction scheme 
is compatible with the known quantum gate-assembly schemes. This is a good improvement.  
\item {\em Efficiency of quantum network}. Because our scheme can compatible with the known scheme, the known decomposition to the array of quantum gates can be directly transformed to our construction, even use it as a quantum subnetwork. Therfore, the efficiency of quantum network in our scheme is not lower than the one in the known scheme. In practice, if one wants to find an effective construction for a quantum (sub)network, in general, his/her needs to use the symmetries in the corresponding transformation matrix and physical features in a given quantum system, while these symmetries and features are usually connected with the elements of this transformation matrix. Based on our scheme related to the matrix elements directly, to find an effective construction should be easier than the known schemes 
without obvious and direct relation to the matrix elements. Furthermore, in our scheme, an effective construction for a quantum subnetwork is simpler since itself is just simpler and it can reduce into more simpler hiberarchy of quantum subnetworks. Moreover, our decomposition principle is to let the lowest hierarchy quantum subnetworks has enough symmetries and physical features which can be directly used for an effective construction as possible. In addition, we can use some 
known and knowing in future effective constructions for some quantum 
subnetworks, for example, quantum network for quantum Fourier 
transformation. This is the 
reasons why our construction scheme have more means do it than the 
known ones. This is a good improvement.
\item{\em Design principle}. From our scheme, it can follow a general design principle for a quantum algorithm and quantum simulating. 
That is, we ought to find such a suitable and optimized decomposition 
that the quantum subnetworks for its components can be effectively and 
easily constructed as possible. The process of decomposition can continue 
until our aim is arrived at. In principle, this is possible if a quantum 
computation can be effectively carried out. In order to do this, we need 
to use the fundamental laws of physics, specially the principles and 
features of quantum mechanics. This is a good improvement.
\item{\em Programmability}. It is worth mentioning, in mathematical, our elements of quantum circuit 
are explicitly and directly related with the natural basis of the matrix 
and matrix elements. The natural basis leads them to become some standard accumulating units as a transformation matrix. Resetting their parameters 
corresponds to rewriting the matrix elements and so their combination will 
form a new transformation matrix. Usually, these elements of quantum 
circuit have been combined to build a quantum (sub-)network in a way with 
the effective construction. Resetting their parameters is just to program 
this quantum (sub-)network \cite{Nielsen}. Therefore, our construction 
provides a possible way to program a quantum (sub-)network. 
This is a potential improvement.         
\item {\em Fault tolerance and error control}. In a quantum network, the different parts (subnetworks) have different requirement and ability of fault tolerance and error control in general. Our scheme allows to design respectively different subnetworks, it is helpful and favorable to fault tolerance and error control. For example, in Sec. \ref{S5}, we discuss a heterotic connection in which one has analyzed the efficiency of some simple error correction techniques for a quantum subnetwork for modular exponentiation. This is a possible improvement. 
\item{\em Industrialization and commercialization}. Our quantum network is uniform and standardized. This is a necessary condition of industrialization. Our quantum network is updatable and repairable. This is a necessary condition of commercialization. Both of them are consist with principle of economy. This is a useful improvement.
\end{enumerate}    

Note that only an effective construction is really useful, the uniform structure,
universality and scalability are then restricted to some content. At present, the fault tolerance and error control are more important problems than the scalable and uniform structure as well as (really) universality. However, with the development of quantum computer, one can not help to consider some problems proposed in this paper. Our scheme may become a concerned model in near future if it is not now. We are sure that our scheme is a possible candidate for a scalable, uniform and universal quantum network. 

Of course, our scheme leaves some open questions. For example, how to program it and how to implement it in experiment. 

This research is on progressing.  

\medskip

I would like to thank Artur Ekert for his great help and for his hosting 
my visit to center of quantum computing in Oxford University. I also 
thank Markus Grassl for his helpful comments.

\end{document}